\documentclass[traditabstract]{aa}

\usepackage{graphicx}
\usepackage{txfonts}
\usepackage{amssymb}

\usepackage{natbib}
\bibpunct{(}{)}{;}{a}{}{,}

\newcommand{\beq}{\begin{equation}}
\newcommand{\eeq}{\end{equation}}
\newcommand{\bea}{\begin{eqnarray}}
\newcommand{\eea}{\end{eqnarray}}
\newcommand{\req}[1]{Eq.~(\ref{#1})}
\newcommand{\dd}{\mathrm{d}} 
\newcommand{\gcc}{\mbox{g cm$^{-3}$}}
\newcommand{\hq}{h_\mathrm{q}}
\newcommand{\kTF}{k_\mathrm{TF}}
\newcommand{\nion}{n_\mathrm{ion}}

\begin{document}

\title{Thermonuclear fusion in dense stars}
\subtitle{Electron screening, conductive cooling,
and magnetic field effects}

\author{A. Y. Potekhin
  \inst{1,2,3}
  \thanks{\email{palex@astro.ioffe.ru}}
\and
G. Chabrier
  \inst{1,4}
  \thanks{\email{chabrier@ens-lyon.fr}}
  }

\institute{CRAL (UMR CNRS No. 5574), 
Ecole Normale Sup\'{e}rieure de Lyon,
69364 Lyon Cedex 07, France
\and
Ioffe Physical-Technical Institute,
Politekhnicheskaya 26, 194021 St.~Petersburg, Russia
\and
Isaac Newton Institute of Chile, 
         St.~Petersburg Branch, Russia
\and
School of Physics, University of Exeter, Exeter, UK EX4 4QL}
\date{Received ... / Accepted ...}

\abstract{We study the plasma correlation effects on
nonresonant thermonuclear reactions of carbon and
oxygen in the interiors of white dwarfs and liquid envelopes
of neutron stars.
We examine the effects of electron screening on thermodynamic
enhancement of thermonuclear reactions in dense plasmas beyond
the linear mixing rule. Using
these improved enhancement factors, we calculate
carbon and oxygen ignition curves in white
dwarfs and neutron stars. The energy balance and ignition
conditions in neutron star envelopes are evaluated,
taking their detailed thermal structure into account. The result
is compared to the simplified
``one-zone model,'' which is routinely used in the literature. We also
consider the effect of strong magnetic fields on 
the ignition curves in the ocean of magnetars.
}

\maketitle

\section{Introduction} 
\label{sect:intro}

Thermonuclear reactions play a crucial role in
stellar evolution scenarios. In particular, they are important in
white dwarfs and neutron stars. Ignition of
degenerate carbon and oxygen in the interiors of white dwarfs
gives rise to Type Ia supernovae
\citep{HoyleFowler60,HillebrandtNiemeyer00}. Nuclear reactions in
degenerate envelopes of accreting neutron stars are responsible
for X-ray bursts and for the overall chemical and thermal
structure of the envelopes \citep{FushikiLamb87,BrownBildsten98}.
Nuclear fusion rates in the interiors of degenerate stars can be
significantly enhanced over the binary \citet{Gamow28} rates
because of the many-body screening effect in the dense plasma
(\citealp{Schatzman48}; for reviews, see \citealp{YaSha} and
\citealp{Ichimaru93}). 

The screening in degenerate matter is usually
treated under the assumption that the electron gas can be
considered as a uniform ``rigid'' background.
The influence of the electron polarization on the enhancement of
nuclear reaction rates has been studied in some detail
in several papers 
\citep{Salpeter54,ItohTI77,YaSha,SahrlingChabrier98,Kitamura00},
which confirms that the electron screening
effect is rather weak in degenerate matter. At the time of these
studies, uncertainties in the reaction rates due to other
factors, viz.\ quantum effects and deviations from the linear
mixing rule (LMR) in strongly coupled plasmas, as well as
theoretical uncertainties in the nuclear effective potentials at
short distances, were more important than the polarizable
electron-screening
effects.

The mentioned uncertainties have been
substantially reduced in recent years. 
There has been significant progress in
treating cross sections of binary nuclear fusion reactions
(\citealp{Beard-ea10}, and references therein).
\citet{Yakovlev-ea10} constructed an analytic model for
calculating these cross sections, which  accurately describes
the data and parametrized it for a
number of C, O, Mg, and Ne isotopes. \citet{PollockMilitzer04}
and \citet{MilitzerPollock05} used the path-integral Monte Carlo
(PIMC) method to determine contact probabilities of
reacting nuclei for one-component plasma (OCP) with emphasis on
many-body quantum effects (these calculations supersede the
previous PIMC study by \citealp{Ogata97}). \citet{ChugunovDWY07}
compared these PIMC results to semiclassical
calculations and find good agreement between the two approaches at
temperatures higher than about one fifth of the ion plasma temperature.
These authors
also obtained a simple parametrization of the reaction
rates with allowance for the ion quantum effects.
\citet{ChugunovDW09a} extended these results to reactions between
different nuclei and suggested an analytic expression for reaction
rates in multicomponent ion mixtures, based on the LMR.
\citet{ChugunovDW09b} used extensive Monte Carlo simulations and
discuss corrections to the LMR for the plasma-screening
function in strongly coupled binary ionic mixtures. They also
propose an analytic formula for the screening function in
ion mixtures. 

\citet{ChugunovDW09a,ChugunovDW09b} have employed the model of rigid
electron background. In this paper we  demonstrate that the
electron screening effects are not negligible compared to the
other improvements considered in recent publications. We
derive a simple analytic formula for a quick evaluation of these
effects. We also calculate the ignition curves for carbon,
oxygen, and their mixtures. We consider plasma cooling
by heat conduction and different neutrino emission mechanisms,
which evacuate the heat released in nuclear burning, thereby
determining the ignition curve. The account of the heat diffusion
is taken by detailed calculation of the thermal structure of
neutron star envelopes and corresponding heat fluxes. The result
is compared to the simplified ``one-zone approximation''
\citep{BrownBildsten98,CummingBildsten01,Gasques_ea07}. Finally,
we consider the effects of strong magnetic fields on the ignition
curves in neutron star envelopes.

In Sect.~\ref{sect:enh} we compare
different approximations for the enhancement factors and
study the effect of electron screening. In Sect.~\ref{sect:ign}
we calculate carbon and oxygen ignition curves in degenerate
stars, using the state-of-the-art treatment of carbon and oxygen
fusion reactions,  neutrino emission mechanisms, and heat
conduction with allowance for strong magnetic fields. Results
are summarized in Sect.~\ref{sect:concl}.

\section{Enhancement factors}
\label{sect:enh}

\subsection{Classical theory and modern approximations}
\label{sect:enh-gen}

It is customary to write the cross section of binary
nuclear fusion reactions in the form (e.g.,
\citealp{Yakovlev-ea10})
\beq
   \sigma(E) = \frac{\mathrm{e}^{-2\pi \eta}}{E}\,S(E),
\eeq
where $E$ is the center-of-mass kinetic energy of the reacting
nuclei ``1'' and ``2'', 
\beq
   \eta=\sqrt{\frac{E_R}{E}},
\qquad
   E_R = \frac{(Z_1 Z_2 e^2)^2\,m_{12}}{2\hbar^2},
\eeq
$Z_j e$ is the charge of nucleus ``$j$'', $e$ is the elementary
charge, $m_{12} = m_1 m_2/(m_1+m_2)$ is the reduced mass, and
$S(E)$ is a function called ``astrophysical factor.''
For the Boltzmann distribution of nuclei,
the reaction rate (the number of fusion events per unit time
in unit volume) in the absence of screening is given by
\beq
   R_{12}=w_{12}\,n_1 n_2
    \left( \frac{8}{\pi m_{12} T^3} \right)^{1/2}
    \int_0^\infty \mathrm{e}^{-2\pi\eta-E/T} S(E)\,\dd E,
\label{R}
\eeq
where $n_j$ is the number density of the ions of type ``$j$'',
$T$ is temperature in energy units, and the factor $w_{12}$ accounts
for statistics: $w_{12}=\frac12$, if nuclei ``1'' and ``2'' are
identical; otherwise $w_{12}=1$. If $T$ is small, then the
integrand in \req{R} is strongly peaked at the energy
$E_\mathrm{pk}=(\pi^2 E_R T^2)^{1/3}$, and the integral can be
evaluated as \citep{SalpeterVH69}
\beq
   \int_0^\infty \mathrm{e}^{-2\pi\eta-E/T}\,S(E)\,\dd E
   \approx
    \left(\frac{4\pi}{3} T E_\mathrm{pk}\right)^{1/2} S(E_\mathrm{pk})
    \,\mathrm{e}^{-\tau},
\label{saddle}
\eeq
where 
\beq
   \tau = 3(\pi^2E_R/T)^{1/3}.
\label{tau}
\eeq
Approximation (\ref{saddle}) is valid, if $\tau\gg1$.

In order to take
the plasma screening effects into account,
it is
convenient to write the radial pair-distribution function for
ions in the form
\beq
   g_{12}(r) = \exp\bigg( - \frac{Z_1 Z_2 e^2}{rT} \bigg)
      \, \exp\bigg(\frac{H_{12}(r)}{T} \bigg) ,
\eeq
where the first factor is the Boltzmann formula for an ideal gas,
while the second one shows how the probability of separation of
two chosen ions is affected by the surrounding plasma particles.
The function $H_{12}(r)$ is often called \emph{screening
potential} of the plasma \citep[e.g.,][]{DeWittGC73}.

Along with the customary ion sphere radii  $a_j=(3Z_j/4\pi
n_e)^{1/3}$, where $n_e$ is the electron number density, and
Coulomb coupling parameters $\Gamma_j=(Z_j e)^2/a_j T$, it is
convenient to introduce parameters
\beq
   \Gamma_{12}=\frac{Z_1 Z_2 e^2}{a_{12}T},
\qquad
   a_{12}=\frac{a_1+a_2}{2} .
\label{Gamma12}
\eeq
Provided that $H_{12}(r)$
varies slowly on the scale of the classical turning point
distance, which requires that $3\Gamma_{12}/\tau\ll1$
\citep{Ichimaru93}, the \emph{screened reaction rate} is
approximately given by
$R_{12}\mathrm{e}^{h}$, where the enhancement exponent is 
 \citep{Salpeter54}
\beq
   h = H_{12}(0)/T,
\label{H0}
\eeq
and $R_{12}$ is given by \req{R} with replacement of $S(E)$ by
$S(E+H_{12}(0))$ \citep{ChugunovDW09a}. As discussed by
\citet{Mitler77} and \citet{ItohTI77},
approximation (\ref{H0}) 
needs to be corrected at higher densities, 
where $3\Gamma_{12}/\tau$
is not small; in the latter case,
the quantum effects on ion motion
become significant \citep{Jancovici77,AlaJanco78}.

The Helmholtz free energy 
$F(V,T;\{N_j\};N_e)$ depends on the numbers $N_j=n_j V$ of ions of
all kinds, the number of electrons $N_e=n_e V$, 
volume $V$ and temperature $T$. We write it
in the form
$F=F_\mathrm{id}+F_\mathrm{ex}$, where 
$F_\mathrm{id}$ is the free energy of the ensemble of
noninteracting ions and electrons, and $F_\mathrm{ex}$ is the
excess free energy that accounts for the interactions.
In this paper we consider only neutral plasmas,
so that $n_e=\sum_j n_j Z_j$. One can rigorously prove
\citep{DeWittGC73,Jancovici77}
that $H_{12}(0)$ equals the difference between the excess free
energies before and after an individual act of fusion.
In the thermodynamic limit this gives the relation
(cf.\ \citealp{IchimaruKitamura96})
\beq
   h = \bigg( \frac{\partial}{\partial n_1} 
     + \frac{\partial}{\partial n_2} 
     - \frac{\partial}{\partial n_3} \bigg)
     \,\left[ \nion f_\mathrm{ex}(\{ n_j \}, n_e ,T) \right],
\label{h0gen}
\eeq
where $\nion=\sum_j n_j$ is the total number density of ions,
including number density $n_3$ of composite nuclei, which have
charge number $Z_3=Z_1+Z_2$ and mass  $m_3\approx m_1+m_2$, and
$f_\mathrm{ex}\equiv F_\mathrm{ex}/\nion VT$ is the normalized
excess energy. 

In strongly coupled Coulomb plasma mixtures of classical ions and
degenerate electrons, the LMR is fulfilled
\citep{HansenVieillefosse76,ChabrierAshcroft90}, so that 
$f_\mathrm{ex}\approx  f_\mathrm{lm}$, where
\beq
   f_\mathrm{lm}(\{ n_j \}, n_e ,T) = \sum_j x_j f_j(n_e,
   T).
\label{LMR}
\eeq
Here, $x_j\equiv n_j/\nion$ denotes the number fractions and
$f_j(n_e, T)$ is the normalized excess free energy
$f_\mathrm{ex}$ for a plasma containing only the $j$th type of
ions. Accurate analytic expressions for $f_j$ in the Coulomb
liquid have been derived in our previous work \citep{PC00},
and we use these expressions hereafter. It follows from
Eqs.~(\ref{h0gen}) and (\ref{LMR}) that the enhancement exponent 
in the LMR approximation is
\beq
   h_\mathrm{lm} = f_1(n_e, T) + f_2(n_e, T) - f_3(n_e, T).
\label{hLMR}
\eeq
In the approximation of rigid electron background, this reduces to
\beq
   h_\mathrm{lm,ii} = f_\mathrm{ii}(\Gamma_1) 
     + f_\mathrm{ii}(\Gamma_2) - f_\mathrm{ii}(\Gamma_3),
\label{hLMRii}
\eeq
where $f_\mathrm{ii}(\Gamma)$ is the normalized excess
free energy of the OCP.
In the ion sphere approximation, 
$f_\mathrm{ii}(\Gamma)=-0.9\,\Gamma$, hence $h_\mathrm{lm,ii}$ 
becomes \citep{Salpeter54}
\beq
   h_\mathrm{S} = 0.9\,(\Gamma_3-\Gamma_1-\Gamma_2).
\label{hS}
\eeq
We note, in passing, that Eq.~(169) of \citet{YaSha}
recovers this equation with
a factor 1.055 instead of 0.9.
We will see, however, that the factor 0.9
provides a much better approximation
to the accurate screening function.

The LMR is not exact, and it becomes progressively
inaccurate with decreasing $\Gamma_j$. When $\Gamma_j\ll1$ for
all $j$, the excess free energy $F_\mathrm{ex}$ is described by the
\citet{DH} approximation, $F_\mathrm{DH} = - VT/12\pi D^3$,
where $D$ is the screening length.
For nondegenerate electrons and ions,
\bea&&
   D^{-2} = D_e^{-2} + D_\mathrm{ion}^{-2},
\label{D}
\\&&
   D_e^{-2} = \frac{4\pi e^2}{T}\, n_e,
\quad
   D_\mathrm{ion}^{-2} = \frac{4\pi e^2}{T}\,\sum_j n_j Z_j^2 .
\eea
In the approximation of rigid electron background, 
$D=D_\mathrm{ion}$. In the Debye-H\"uckel approximation,
\req{h0gen} yields for the enhancement exponent
\citep{Salpeter54}
\beq
   h_\mathrm{DH} = \frac{Z_1 Z_2 e^2}{DT}.
\label{hDH}
\eeq
For an arbitrary degree of degeneracy 
(but at not too strong Coulomb coupling; see \citealp{Chabrier90}), 
the screened interaction between ions is approximately
described by a Yukawa potential, 
$Z_1 Z_2 e^2 \mathrm{e}^{-r/D}/r$, 
with 
$D=\kTF^{-1}$ for electron screening or
$D = \left(\kTF^2 + D_\mathrm{ion}^{-2}\right)^{-1/2}$
for electron and ion screening.
Here,
\beq
 \kTF^2 = 4\pi e^2\, {\partial n_e / \partial \mu_e}
\label{k_TF}
\eeq
is the Thomas-Fermi wave number, and $\mu_e$ is the chemical
potential of Fermi gas of  electrons.
Using Eq.~(24) of \citet{CP98}, one can write $\kTF(n_e,T)$ in
analytic form. We note, however, that the Yukawa model corresponds to
the Thomas-Fermi limit, $\epsilon(k)\sim 1+(\kTF/k)^{2}$, for the
static dielectric function $\epsilon(k)$, which may only be justified
at $k\ll\kTF$ \citep[see, e.g.,][]{GalamHansen76}. Therefore,
this model is inappropriate at short distances (i.e., large
wavenumbers $k$). In particular, it is not applicable for the
evaluation of the screening potential at zero separation,
$H_{12}(0)$.

For the general case, \citet{SalpeterVH69} proposed the following
interpolation between the Debye-H\"uckel and strong-coupling
limits:
\beq
   h_\mathrm{SVH} = \frac{ h_\mathrm{S}\,h_\mathrm{DH}
      }{ 
      \sqrt{h_\mathrm{S}^2 + h_\mathrm{DH}^2}},
\label{SVH}
\eeq
where $h_\mathrm{S}$ and $h_\mathrm{DH}$ are
given by Eqs.~(\ref{hS})\,--\,(\ref{hDH}).

Another analytic approximation for the enhancement factor
beyond the LMR was constructed by \citet{ChugunovDW09b}, based on
Monte Carlo simulation results for the \emph{rigid background} model.

These analytic approximations can be compared to the 
result given exactly by \req{h0gen}. We write the normalized excess
free energy in the form $f_\mathrm{ex} = f_\mathrm{lm} +
f_\mathrm{mix}$, where $f_\mathrm{lm}$ is given by \req{LMR}, and
$f_\mathrm{mix}(\{x_j\},\{Z_j\};n_e,T)$ is the correction to the
LMR, which was recently obtained in analytic form
\citep{Potekhin-ea09}. Then, from \req{h0gen}, we obtain
the enhancement exponent
\beq
   h_0 = h_\mathrm{lm}
    + \left.\frac{\dd
     f_\mathrm{mix}(x_1+\xi,x_2+\xi,x_3-\xi)
      }{
      \dd\xi} \right|_{\xi=0},
\label{hmix}
\eeq
where $h_\mathrm{lm}$ is given by \req{hLMR}. 
Figure~\ref{fig:enhclass} shows enhancement
factors for $^{12}$C fusion in different approximations,
normalized to the \citet{SalpeterVH69} 
enhancement factor approximation (\ref{SVH}). Here,
we intentionally neglect ion quantum effects and postpone
their discussion to Sect.~\ref{sect:quant}.
We compare
the analytic expressions for the OCP (thus rigid background: 
the fit of \citealt{ChugunovDW09b}, 
and the result of using \req{hmix}
for a rigid background, i.e. with $h_\mathrm{lm}$
replaced by  $h_\mathrm{lm,ii}$ and
$f_\mathrm{mix}$ given by the fit of \citet{Potekhin-ea09}
 for the rigid background case.
Hereafter, this approximation 
will be denoted $h_\mathrm{ii}$.
Comparison of the two dot-dashed curves shows that these
approximations agree with each other
within typically 2\%.

\subsection{Electron screening}
\label{sect:enh-el}

\begin{figure}
\includegraphics[width=\columnwidth]{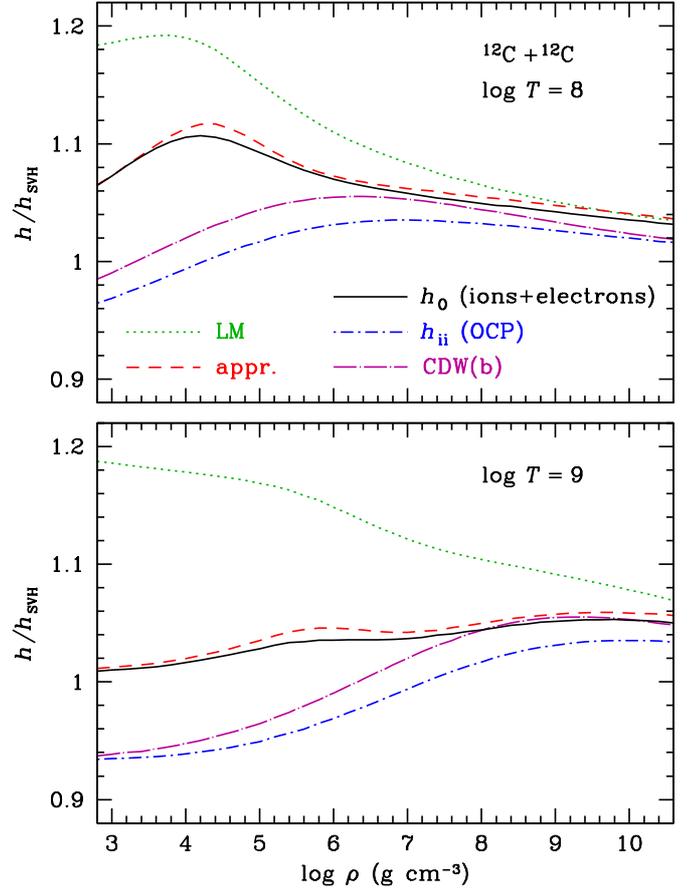}
\caption{Plasma enhancement exponents for carbon fusion reactions
in different approximations, neglecting the ion quantum effects,
normalized to the enhancement exponent
given by \req{SVH}, as functions of mass density
for 2 isotherms (top panel: $T=10^8$~K; 
bottom panel: $T=10^9$~K). Dot-dashed lines
correspond to the electron rigid background model. Namely,
dot-short-dashed lines show $h_\mathrm{ii}$ given by
\req{hmix} for a rigid background, while 
dot-long-dashed lines show the approximation of \citet{ChugunovDW09b}. 
The other lines correspond to the polarizable electron
background case: dotted line: LMR [\req{hLMR}];
solid line: $h_0$ [\req{hmix}]; dashed line:
$h_\mathrm{appr}$ [\req{appr}].
}
\label{fig:enhclass}
\end{figure}

In Fig.~\ref{fig:enhclass} we compare the enhancement factors
obtained using
\req{hmix}, where $f_\mathrm{mix}$ is given by the fit of 
\citet{Potekhin-ea09}, for the case of a
\emph{polarizable electron} background, and for the OCP.
This comparison
illustrates the contribution of electron gas polarization to the
screening exponent under the present conditions. Additionally
we show $h_\mathrm{lm}$, given by
\req{hLMR}, with the electron polarization taken into account in
$f_j(n_e,T)$. We see that the correct enhancement factor
differs appreciably from the LMR result in the
low-density regime, and the difference increases with
temperature.

The electron screening contribution can be quickly estimated as
follows. We define the effective ion screening
length for the reacting nuclei as
\beq
   \widetilde{D}_\mathrm{ion} = 
   \sqrt{D_\mathrm{ion}^2 + (0.6\,a_{12})^2}.
\eeq
At low densities, where the Debye-H\"uckel theory
is appropriate, $\widetilde{D}_\mathrm{ion}$ approaches
$D_\mathrm{ion}$ while at high densities it is proportional to
$a_{12}$. The numerical factor 0.6 is the only fitting parameter.
Then the approximation for $h_0$ reads as
\beq
   h_\mathrm{appr} = h_\mathrm{ii}
   \sqrt{1+\kTF^2\,\widetilde{D}_\mathrm{ion}^2}.
\label{appr}
\eeq
The result is also illustrated in Fig.~\ref{fig:enhclass}.

\subsection{Quantum effects}
\label{sect:quant}

\begin{figure}
\includegraphics[width=\columnwidth]{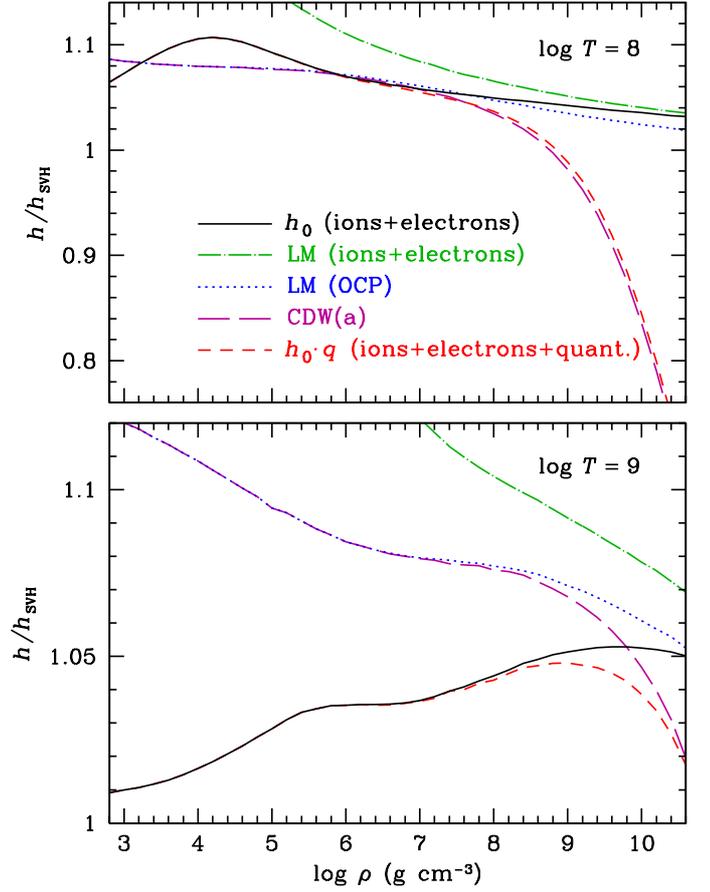}
\caption{Same enhancement exponents as in
Fig.~\ref{fig:enhclass}, but with other approximations. Solid,
dotted, and dot-dashed
lines correspond to results while neglecting ionic quantum
effects on the enhancement factor: the dotted line
corresponds to the LMR for OCP [\req{hLMRii}], the dot-dashed line
demonstrates LMR for the polarizable electron background
[\req{hLMR}],
and the solid line
shows $h_0$ beyond the LMR approximation
[\req{hmix}]. The dashed lines illustrate the  impact
of ionic quantum effects: the long-dashed line presents
the fit of \citet{ChugunovDW09a}, while the short-dashed line is
the approximation $q h_0$,
which includes both the ionic and electronic screening
contributions and takes both the quantum effects 
and the deviations from the LMR into account
(see text).
}
\label{fig:enhquant}
\end{figure}

\citet{Jancovici77} and \citet{AlaJanco78} examined the
short-range behavior of the internuclear correlation functions
and  showed that the quantum effects for the ions
decrease the enhancement
factor, which we write as $\mathrm{e}^{\hq}$. 
They developed a perturbation
expansion of the enhancement exponent $\hq$ in powers of the
parameter $(3\Gamma/\tau)$. This theory is
applicable at $\Gamma\gg1$ and $3\Gamma/\tau\lesssim1$.
\citet{PollockMilitzer04} and \citet{MilitzerPollock05} performed
PIMC calculations of the contact probabilities in the quantum
regime. They confirm the conclusions of
\citet{Jancovici77} and \citet{AlaJanco78} and extended numerical
results beyond the applicability range of the perturbation
theory.  \citet{ChugunovDWY07} find that the results of
\citet{MilitzerPollock05} agree with semiclassical calculations
and suggest an analytic parametrization of the reaction rates
that accounts for the quantum effects in an OCP. A similar
parametrization for multicomponent mixtures has recently been
derived by \citet{ChugunovDW09a}, in the LMR
approximation. They find that the quantum
effects can be described by the use of
\req{hLMRii} with the substitution\footnote{Here 
a typo in \citet{ChugunovDW09a} is corrected. We are
grateful to A.I. Chugunov for drawing our attention to
another typo in the first version of this our paper.}
$\Gamma_j \to \tilde\Gamma_j = \Gamma_j/t_{12}$, where
\bea&&\!\!
   t_{12}=\left[1+c_1\, (3\Gamma_{12}/\tau)
    + c_2\, (3\Gamma_{12}/\tau)^2
     + c_3\, (3\Gamma_{12}/\tau)^3 \right]^{1/3},
\\&&\!\!
  c_1 = 0.013\,\gamma^2,
\quad
  c_2 = 0.406\,\gamma^{0.14},
\quad
  c_3 = 0.062\,\gamma^{0.19}+1.8/\Gamma_{12},
\nonumber
\eea
$\tau$ and $\Gamma_{12}$ are defined by Eqs.~(\ref{tau}) and 
(\ref{Gamma12}), and $\gamma=4Z_1 Z_2 / (Z_1 + Z_2)^2$
(see Fig.~\ref{fig:enhquant}).

Expansions of the fitting functions of \citet{ChugunovDWY07} and
\citet{ChugunovDW09a} in Taylor series do not recover the
perturbation series of \citet{AlaJanco78}. This mismatch, however, is
probably unimportant (unless one is interested in the second and
higher derivatives of $h$), because the numerical agreement with
the Alastuey-Jancovici results in their validity domain 
($3\Gamma/\tau < 1$) is quite good.

To include the quantum
effects in the general case, we multiply the classical
expression (e.g., Eqs.~(\ref{hmix}) or (\req{appr})) 
by the quantum decreasing factor
$q=\tilde{h}_\mathrm{lm,ii}/h_\mathrm{lm,ii}$, where
$h_\mathrm{lm,ii}$ is given by \req{hLMRii}, and
$\tilde{h}_\mathrm{lm,ii}$ results from the 
replacement of $\Gamma_j$ by $\tilde{\Gamma}_j$ in \req{hLMRii}.
The function $h_0\cdot q$ is also plotted in Fig.~\ref{fig:enhquant}. 

\subsection{Discussion}

Figures \ref{fig:enhclass} and \ref{fig:enhquant} demonstrate
that electron screening always increases the value of the
enhancement factor.
This result is intuitively expected, because
allowance for additional screening particles augments the overall
effect.  It qualitatively agrees also with the findings of
\citet{SalpeterVH69}, \citet{DeWittGC73}, \citet{YaSha},
 \citet{SahrlingChabrier98}, and \citet{Kitamura00}.
The opposite result was claimed by \citet{PollockMilitzer04},
who found that electron screening ``reduces the enhancement
effect.'' This confusion arises from their use of a Yukawa
potential to describe the electron screening. We pointed out in
Sect.~\ref{sect:enh-gen} that the Yukawa model becomes incorrect at short
distances; in particular, it is incapable of determining the contact
probabilities between fusing nuclei. 
\citet{Ichimaru93} mentioned two opposite effects
of electron screening: first, the binary repulsive potentials
between reacting nuclei are reduced by electrons (``short-range
effect''), which increases $H_{12}(0)$; second, the reduction of
particle interactions by the screening affects the many-body
correlation function in such a way that it decreases $H_{12}(0)$ 
(``long-range effect''). In
real electron-ion plasmas (without the Yukawa approximation) the
first effect overpowers the second one. The Yukawa model grasps
the second effect, but misses the first, dominant one.

In the high-density domain ($\rho\gtrsim10^8$ \gcc), where the
fusion ignition in dense stars occurs, both corrections to $h$
due to electron screening and due to the
deviations from the LMR (see Fig.~\ref{fig:enhclass}) 
become quite small, 
of a few percent or less. Moreover, since these two
corrections have opposite signs, they can balance each other out.
This is illustrated in Fig.~\ref{fig:enhquant}, where 
the accurate result is the same as in
Fig.~\ref{fig:enhclass}, and we show also
the result of application of the LMR
for the cases of polarizable electron background according to
\req{hLMR} and the rigid background with LMR approximation
according to \req{hLMRii}.

At lower density, the corrections to the enhancement
exponent $h$ due to electron polarization and deviation from the
LMR become relatively large, but $h$ itself becomes
rather small, so that its effect on the reaction rate
is not very significant. This domain concerns
partially degenerate objects, such as low-mass stars or brown
dwarfs, and may affect the nuclear production of light elements,
such as deuterium, lithium, beryllium, etc., which provide
tracers for the mass and/or age determination of these objects
\citep{ChabrierBaraffe00}. These situations will be examined in
a forthcoming paper.

One can note that the simple approximation of
\citet{SalpeterVH69}, which we have chosen for
normalization, performs surprisingly well: in the  presently
explored $\rho$\,--\,$T$ domain of astrophysical interest,
it provides an accuracy of the enhancement factor
better than 10\%, i.e., better than $0.04$ for
$\log_{10}\sigma(E)$. Some later approximations
\citep{YaSha,ItohKM90} are not as good. (They would fall
outside the frames of Fig.~\ref{fig:enhquant}). 

Since the uncertainties in the nuclear part of the
reaction rates are still larger (see, e.g.,
\citealt{Aguilera-ea06}), in practice it appears sufficient
to use the approximation of \citet{SalpeterVH69} with the
quantum correction:
\beq
   h\approx h_\mathrm{SVH}\, q,
\label{happr}
\eeq
where $h_\mathrm{SVH}$ and $q$ are given in
Sects.\,\ref{sect:enh-gen} and \ref{sect:quant},
respectively. Including further improvements to $h$
is currently just a question of completeness. However, as we
have seen in Sect.\,\ref{sect:enh-el}, the
electron-polarization corrections are generally comparable
to or even larger than other corrections discussed in
literature. Therefore, the electron screening should
be taken into account in every treatment of the
thermodynamic enhancement factor that goes beyond
approximation (\ref{happr}).

\section{Ignition curves}
\label{sect:ign}

\subsection{Nuclear heating and neutrino cooling}
\label{sect:ign-neu}

The ignition curve is the line in the $\rho$\,--\,$T$ plane  that
determines the highest densities and temperatures at which 
exothermic nuclear reactions in the plasma can be stable against
thermal runaway. It is determined by the balance between nuclear
energy generation rate and local heat losses. In this subsection
we focus on the case where the heat losses are mainly caused
by neutrino emission,  which is appropriate in white dwarfs,
e.g., for modeling supernova Ia events
\citep{HillebrandtNiemeyer00}. 

The thermonuclear fusion rate is given by \req{R}, where for
$S(E)$ we substitute the parametrization of
\citet{Yakovlev-ea10}. The energy release power per unit volume
equals $R_{12}Q_{12}$, where
$Q_{12}$ is the energy release in a single fusion
event. We use the values of $Q_{12}$ given by
\citet{FowlerCZ75}:
13.931~MeV for $^{12}\mathrm{C} + {^{12}\mathrm{C}}$,
16.754~MeV for $^{12}\mathrm{C} + {^{16}\mathrm{O}}$, and
16.541~MeV for $^{16}\mathrm{O} + {^{16}\mathrm{O}}$
reactions. 

At very high densities, pycnonuclear fusion due to zero-point ion
vibrations \citep{Cameron59} becomes more important than the
thermonuclear one. We calculate it 
using Eq.~(33) and
Table~II (the first line) of \citet{Yakovlev-ea06} and add
it to the thermonuclear rate. 
The ignition curves portrayed in the
figures, however, never enter the domain of pycnonuclear
burning.

The dominant mechanism of energy loss due to neutrino
emission differs, depending on the physical conditions. For the
present conditions of interest, the main mechanisms are
neutrino bremsstrahlung by electron scattering off nuclei,
plasmon decay, and electron-positron annihilation. The energy
loss rates are given, respectively, by Eqs.~(76), (38), and (22)
of \citet{Yakovlev-ea01}.

\begin{figure}
\includegraphics[width=\columnwidth]{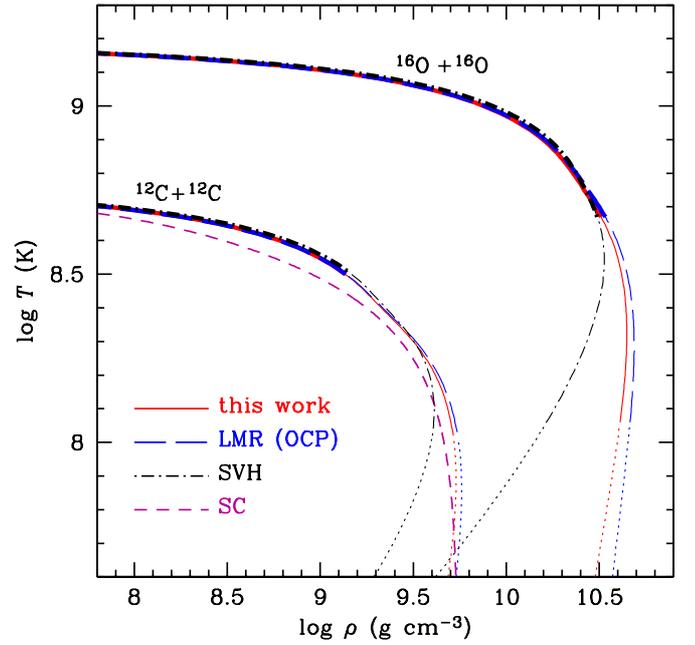}
\caption{Ignition curves for carbon (lower/left group of lines)
and oxygen (upper/right group of lines) in different
approximations. Solid lines are obtained using \req{hmix}
and the quantum correction according to Sect.~\ref{sect:quant};
long-dashed lines: \req{hLMRii} and the quantum correction; 
dot-dashed lines: \req{SVH} without any correction.
The lines are plotted heavy if the characteristic
nuclear fuel burning time $t_\mathrm{burn}$ is shorter than
1~Myr; the lines are plotted thin if
$\mbox{1~Myr} <t_\mathrm{burn}<14$~Gyr. In the domain where
$t_\mathrm{burn}>14$~Gyr the lines are dotted. For
the carbon case we also plot the fit \citep{Potekhin-ea03} to the
results of \citet{SahrlingChabrier98} (the short-dashed line).
}
\label{fig:ignco}
\end{figure}

Figure \ref{fig:ignco} illustrates the ignition curves for
$^{12}$C+$^{12}$C and $^{16}$O+$^{16}$O reactions in different
approximations. Here, the ignition curves are obtained 
with the enhancement factors given by \req{hmix} with the
quantum correction included according to Sect.~\ref{sect:quant},
and are copmpared with the approximations of
\citet{ChugunovDW09a} and \citet{SalpeterVH69}. 
In the low-temperature region, where
the characteristic burning time $t_\mathrm{burn}$ that
is required to consume 63\% of the nuclear fuel
exceeds the Universe age,
the ignition curves lose any astrophysical sense, because the
burning becomes unrealistically slow, and also because the
poorly known quantum effects become too strong (see the
discussion in \citealp{Gasques-ea05,Yakovlev-ea06}). 

We see that the electron screening slightly shifts the
ignition curves to lower densities, a consequence of the
increased enhancement factor, as mentioned previously. This
result agrees with the previous findings by
\citet{SahrlingChabrier98} and \citet{Kitamura00}.%
\footnote{The carbon ignition curve of
\citet{Kitamura00} substantially differs from our results,
because it was calculated under the assumption of a fixed
neutrino emission power.} In the figure we have also
plotted the carbon ignition curve of
\citet{SahrlingChabrier98} as fitted by
\citet{Potekhin-ea03}, which is close to our current result.
The difference is mainly caused by modern improvements
in the astrophysical factor \citep{Yakovlev-ea10} and
neutrino reaction rates \citep{Yakovlev-ea01}, included in
the present treatment. 
In Appendix~\ref{sect:ignfit} we present a fit to
the current carbon and oxygen ignition
curves in a wide density range.

\subsection{Conductive cooling}

Nuclear burning in neutron stars occurs in relatively thin 
envelopes, whose geometric depth is not more than a few percent
of the stellar radius (e.g., \citealp{BrownBildsten98}, and
references therein). Thermal conductivity in these envelopes is
high and thermal relaxation time is short, so that cooling by
heat diffusion can stabilize nuclear burning beyond the ignition
limits considered in Sect.~\ref{sect:ign-neu}. 

\citet{HansenVanHorn75} estimated stability of H and He burning
in envelopes of neutron stars by comparison of  the
characteristic time scale of thermonuclear heating with
characteristic time scales for removal of energy from the shell
by radiative and conductive thermal diffusion.
\citet{FushikiLamb87} introduced a
differential criterion for such estimates. 
They defined the boundary of thermal
stability of the nuclear burning from the condition
\beq 
   \frac{\delta \epsilon_\mathrm{nuc}(y,T)}{\delta T}
   =
   \frac{\delta \epsilon_\mathrm{cool}(y,T)}{\delta T},
\label{FushikiLamb}
\eeq
where $y$ is the column depth of the burning material (carbon or
oxygen in our case), $\delta T$ is a variation in temperature,
$\delta \epsilon_\mathrm{nuc}$ 
is the respective variation in the
nuclear energy release rate per unit mass
($\epsilon_\mathrm{nuc}=R_{12}\,Q_{12}/\rho$),
and $\delta \epsilon_\mathrm{cool}$ is the variation
of the cooling rate $\epsilon_\mathrm{cool}=-\dd F_r(y)/\dd y$,
where $F_r$ is the outward radial heat flux through unit surface. The
cooling rate $\epsilon_\mathrm{cool}$ depends on the first and
second derivatives of the temperature, therefore the condition
(\ref{FushikiLamb}) is not local, but depends on the overall
temperature and density structure of the envelope. 

Thermal relaxation of the neutron star crust occurs on the scale
of a few tens of years, while the relaxation time of the outer
envelopes is still shorter \citep{Lattimer-ea94,GYP,Fortin-ea10}.
Therefore, the outer envelopes are considered quasistationary for
most of the astrophysical problems, and their thermal structure
is calculated at stationary equilibrium \citep{GPE}. In
particular, \citet{FushikiLamb87} applied this approximation to
the stability analysis of H and He shells of accreting neutron
stars, and \citet{BrownBildsten98} and \citet{CummingBildsten01}
applied it to the stability analysis of carbon shells. The
problem is that \req{FushikiLamb} depends not only on the
equilibrium quasistationary temperature profile $T(y)$, but also
on the profile of its variation $\delta T(y)$. In equilibrium
$\dd F_r(y)/\dd y = \epsilon_\nu - \epsilon_\mathrm{nuc} + T
\partial s/\partial t$, where $\epsilon_\nu$ is
the neutrino emission power per unit mass,
$s$ is the entropy per unit mass,
and $t$ is time. (Here for simplicity we neglect the General
Relativity corrections; for the complete set of accurate
equations see, e.g., \citealp{Richardson-ea79}.) In the
quasistationary approximation, one neglects  $\partial s/\partial
t$. As a result, were the variations $\delta T(y)$ in equilibrium, one
would have
$\epsilon_\mathrm{cool}=\epsilon_\mathrm{nuc}-\epsilon_\nu$,
which exhausts \req{FushikiLamb}. Meanwhile, different
nonequilibrium forms of $\delta T(y)$ lead to different values of
$\delta \epsilon_\mathrm{cool}$. \citet{FushikiLamb87} assumed
the global temperature variation, $\dd \delta T/\dd y=0$. For
example, an assumption of the overall variation in the internal
energy would result in $\dd \delta T/\dd y = -(\delta T/c_P)\,\dd
c_P/\dd y$, where $c_P(y)$ is the heat capacity per unit mass at
constant pressure, taken at equilibrium.

In modern literature
\citep{BrownBildsten98,CummingBildsten01,Gasques_ea07}, a
simplified ``one-zone approximation'' is used for the cooling rate in
\req{FushikiLamb}:
\beq
   \epsilon_\mathrm{cool} = \rho\,\kappa \,T/\tilde{y}^2,
\label{1zone}
\eeq
where $\kappa$ is thermal conductivity,
$\tilde{y}=P/g$ is the column depth
in the plane-parallel nonrelativistic approximation, $P$
is pressure, and $g$ is the surface gravity. 
Definition (\ref{1zone}) is
local and, therefore, free of uncertainties. Because it is local,
the variational derivatives $\delta\epsilon/\delta T$ in
\req{FushikiLamb} can be replaced by the partial derivatives
$\partial\epsilon/\partial T$. However, it misses the information
about the real dependence of conductivity on column depth, which
is fraught with risk being inaccurate.

\begin{figure}
\includegraphics[width=\columnwidth]{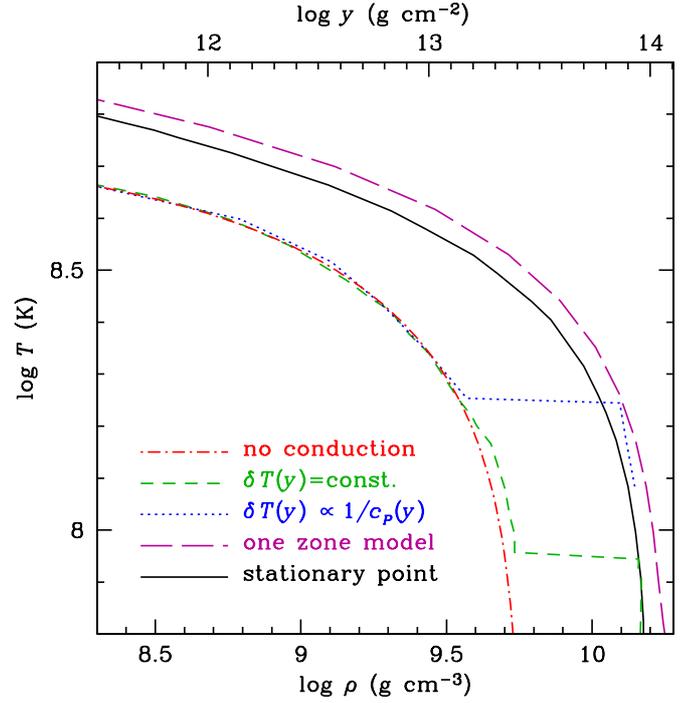}
\caption{Carbon ignition curves in 
the ocean of a neutron star with mass $M=1.4\,M_\odot$
and radius 12 km, according to different models: solid line
corresponds to
\req{stat}, long-dashed line to \req{1zone},
short-dashed and dotted lines, respectively, to the original and
modified approximations of \citet{FushikiLamb87},
\req{FushikiLamb}. For
comparison, the ignition curve without account of conductive
cooling is plotted by the dot-dashed line.
}
\label{fig:pwcb0}
\end{figure}

We define the ignition curve in the outer envelope of neutron
stars in the quasistationary approximation, taking into account
the detailed thermal and mechanical structure of the envelope and
assuming equilibrium variations of the temperature profile
$\delta T(y)$. For definiteness, let us suppose that $\delta
T>0$. For a surface element $\dd S$ of a shell confined between $y_1=y$
and $y_2=y+\dd y$, an increase in temperature $\delta T$ leads to
an increase in the nuclear energy release power
$\delta\epsilon_\mathrm{nuc}\,\dd y\dd S$, an increase in neutrino
energy emission power $\delta\epsilon_\nu\,\dd y\dd S$, and changes
of the heat flux through the outer and inner boundaries of the
shell, $\delta F_{1,2}\, \dd S = \delta F_r(y_{1,2})\, \dd S$. If 
$\delta\epsilon_\mathrm{nuc}\leqslant\delta\epsilon_\nu$,  the
increase in the nuclear heat release is compensated for by the
increase in the energy emission, and the nuclear burning is
stable. 

We now consider the case where 
$\delta\epsilon_\mathrm{nuc} > \delta\epsilon_\nu$. If  the
stationary equilibrium requires an increase in the heat income
rate $F_2\,\dd S$
through the inner boundary, i.e., $\delta F_2 > 0$, 
it means that the increase in the net heat release
in the considered volume,
$\delta\epsilon_\mathrm{tot} =  \delta\epsilon_\mathrm{nuc} -
\delta\epsilon_\nu$, is overcompensated for by the increase in the
outward flux to the stellar surface. In this case, the nuclear
burning is stable. In the opposite case, where $\delta F_2 < 0$, 
$\delta\epsilon_\mathrm{tot}$ is not compensated by the
conductive energy escape to the surface, so that thermal runaway
occurs. Therefore, the largest column depth $y$, at which the
burning can be stable, corresponds to the condition
\beq
   \delta F_r(y) / \delta T(y) = 0. 
\label{stat}
\eeq
At the column depth where the condition (\ref{stat}) is
satisfied, the net energy release $\delta\epsilon_\mathrm{tot}$
is balanced exactly by the increase in surface luminosity.
Beyond this stationary point, thermal runaway starts.

Figure~\ref{fig:pwcb0} illustrates different approximations for
the carbon ignition curve in the ocean of a typical neutron star
with mass $M=1.4\,M_\odot$ and radius 12 km. 
The carbon ignition curve without conductive cooling
is calculated according to \req{ignfit}. 
The
other lines have been obtained by calculating series of
temperature profiles for carbon envelopes of the star, assuming
different surface luminosities and applying different ignition
conditions. The thermal structure has been calculated using the
same code as in \citet{Kaminker-ea09}, but  the artificial
heating model of \citeauthor{Kaminker-ea09} has been replaced by the
nuclear heating. The line corresponding to the
\citet{FushikiLamb87} model shows a jump at a certain
temperature, which corresponds to a switch from conductive to
neutrino cooling. The switch signifies that at higher
temperatures the constant variation $\delta T$ cannot provide a
powerful enough off-equilibrium cooling rate
$\delta\epsilon_\mathrm{cool}$ to compete with the neutrino
cooling. Another functional choice of $\delta T(y)$,
corresponding to a constant variation of internal energy per
mass, is shown by the dotted line. It displays a similar switch
at a higher temperature, therefore the conductive cooling
provided by this variation appears more efficient. The
underestimation of the ignition densities and temperatures by the
two artificial forms of $\delta T(y)$
above the jump temperatures indicates that
these functional variations are unstable on the characteristic
cooling timescale.

The line corresponding to the
one-zone approximation (\ref{1zone}) is compared to
the positions of the stationary points (\ref{stat}). The latter
two models qualitatively agree with each other, with the quantitative
difference caused by the local approximation embedded in
the one-zone model. 

A similar comparison of the models for
the oxygen ignition curve gives similar results. 
For typical accreting neutron stars with low magnetic fields
and effective temperatures from one to several MK, the
ignition curves obtained in our calculations  using the
``stationary point'' condition (\ref{stat}) can be
reproduced by a \emph{modified} one-zone approximation,
where the right-hand side of \req{1zone} is multiplied by a
constant factor $\alpha$ (or equivalently, an
\emph{effective} thermal conductivity
$\kappa_\mathrm{eff}=\alpha\kappa$ is substituted). As the
surface gravity varies from $g\approx0.7\times10^{14}$
cm~s$^{-2}$ (e.g., for stellar mass $M=M_\odot$ and radius
about 15 km) to  $g\approx4\times10^{14}$ cm~s$^{-2}$ 
(e.g., for $M=2\,M_\odot$ and radius of 10 km), the
correction factor $\alpha$ varies from 0.18 to 0.16 in the
case of carbon shell and from 0.33 to 0.28 in the case of
oxygen shell burning.

\subsection{The effect of strong magnetic field}

A strong magnetic field can affect conductivities and make the
heat transport anisotropic, if the Hall parameter (the product of
electron gyrofrequency and effective relaxation time) is larger
than one (e.g., \citealp{UrpinYakovlev80}, and references
therein). Therefore, it can affect the conductive cooling rate and
shift the ignition curves in neutron star envelopes. We have
evaluated the magnitude of this effect by calculating the
temperature profiles for different magnetic field strengths $B$
and inclinations $\theta_B$ at the neutron star surface. 

\begin{figure}
\includegraphics[width=\columnwidth]{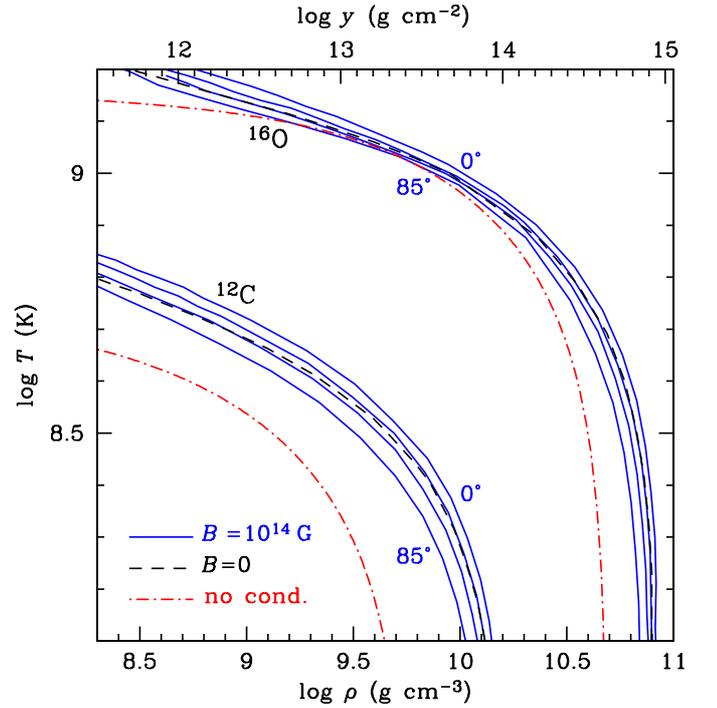}
\caption{Carbon (lower/left group of lines) and oxygen (upper/right group of lines) ignition curves in  the ocean of a
magnetar with mass $M=1.4\,M_\odot$, radius 12 km, and
magnetic field $B=10^{14}$~G for different angles between the
field lines and the normal to the surface: $\theta_B=0^\circ$,
$60^\circ$, $75^\circ$, and $85^\circ$ (respectively, 4 solid
lines from top to bottom in each of the two groups).  For
comparison, the ignition curves given by \req{ignfit} are plotted
by the dot-dashed lines,  and the ignition curves without accounting
for the magnetic field effects are shown by dashed lines.
}
\label{fig:pwb14}
\end{figure}

In strongly magnetized neutron star envelopes, synchrotron
neutrino emission becomes an important energy sink. We calculate
it according to Eq.~(56) of \citet{Yakovlev-ea01}.
Figure~\ref{fig:pwb14} illustrates a few examples. Here,  a
neutron star with $M=1.4\,M_\odot$ and radius 12 km
is supposed to possess a magnetic field $B=10^{14}$~G, typical of
magnetars \citep{Mereghetti08}. The lower lines are drawn
for the carbon envelope, and the upper lines
for the case of oxygen. In each of these groups of lines, the four
solid curves (from top to bottom) are plotted for angles
$\theta_B$ between the field lines and the normal to the surface
equal respectively to $0^\circ$, $60^\circ$, $75^\circ$, and
$85^\circ$. We do not show the case of a strictly tangential field,
because, for any reasonable distribution of the magnetic field
over the stellar surface, the heat transport becomes essentially
two-dimensional around the surface spots where $\theta_B$ is
close to $90^\circ$,
which invalidates the employed approximation of radial
(one-dimensional) transport.
For comparison, we show 
the nonmagnetic ignition curves
and the fusion-vs-neutrino balance curves
discussed in Sect.~\ref{sect:ign-neu} (presumably appropriate for
white dwarfs). 

When the field lines are nearly tangential to the
surface, the strong magnetic field increases the conductive
opacity, because the heat transport perpendicular to the field is
strongly suppressed. Such a field lowers the ignition curves and
thus reduces the stability region. 

In the opposite case of field lines perpendicular to the
surface, the strong magnetic field somewhat raises the ignition
curves and expands the stability region. The latter result could
not be obtained in the one-zone approximation (\ref{1zone}), 
which relies on the local values of thermal conductivity,
$\kappa$, and thermodynamic quantities $\rho$, $T$, and $P$. 
Indeed, in the $\rho$\,--\,$T$ domain shown in
Fig.~\ref{fig:pwb14}, thermal conductivity along the magnetic
field is the same as without the field, because the field is
nonquantizing at $\rho\gg\rho_B$, where, for carbon-oxygen
compositions, $\rho_B=1.41\times10^7\,(B/10^{14}\mbox{~G})^{3/2}$
{\gcc} (e.g., \citealp{P99}). The
equation of state also does not depend on
$B$, if $\rho\gg\rho_B$. Finally, the synchrotron neutrino
emission cannot produce the observed effect, because the neutrino
synchrotron power is less than 5\% of thermonuclear
power along the full ignition lines 
for the conditions appropriate to Fig.~\ref{fig:pwb14}. The
effect is caused by the opacity decrease in the heat-blanketing
envelope at
densities
$\rho\lesssim\rho_B$, where the field is strongly quantizing. The
opacity reduction in this outer region makes the entire
envelope more transparent and  facilitates the heat escape to the
surface. Thus, the
effect is intrinsically nonlocal.

\section{Conclusions} 
\label{sect:concl}

We have studied the effects of electron screening on
thermonuclear reactions in dense plasmas and compared different
approximations to determine plasma enhancement factors for the
nuclear fusion rates. The electron screening always increases the
enhancement effect. The opposite conclusion, sometimes
encountered in the literature, comes from using the
Yukawa potential model, which is inappropriate to calculating the
contact probability for fusing nuclei. The electron screening
correction, which we calculate using accurate analytic
expressions for the free energy, can be satisfactorily
described by a simple approximation (\ref{appr}). This
correction, although small for the case of dense stars, has the
same order of magnitude as other recently suggested corrections
to the enhancement factor, so it needs to be properly
calculated.

Using our analytic formulae for the enhancement factors and
state-of-the-art astrophysical factors for thermonuclear cross
sections, we calculated carbon and oxygen ignition curves in
degenerate stars. The ignition conditions in neutron star
envelopes were evaluated, taking their detailed
thermal structure into account. Comparison of the results to customary
simplified models demonstrates the restricted applicability of
these latter. 

We also studied the effects of strong magnetic fields on the
ignition curves in neutron star envelopes. These results
show that the ignition surface shifts to lower densities in the
stellar regions where the magnetic field is strongly inclined
and to slightly higher densities in the regions of nearly radial
magnetic field.  The latter effect could not be obtained in the
simplified one-zone model. For a magnetar,
the shift in the ignition curve can be
similar to the difference between
the accurate and one-zone calculations
and larger than the correction due to the electron
screening in the plasma. 

\begin{acknowledgements}
A.Y.P acknowledges
useful discussions with Dima Yakovlev and 
partial support from the RFBR Grant 11-02-00253-a
and the Russian Leading Scientific Schools program
(Grant NSh-3769.2010.2).
\end{acknowledgements}
 
\appendix
\section{Fit for C and O ignition curves}
\label{sect:ignfit}

White dwarfs contain a mixture of carbon and oxygen nuclei.
Therefore, it is of practical interest to determine ignition
curves for such mixtures with different number fractions of
carbon ($x_\mathrm{C}$) and oxygen
($x_\mathrm{O}=1-x_\mathrm{C}$).
We have derived a fitting
formula for ignition temperature as a function of mass density,
which is relevant for densities $\rho>100$ \gcc,
temperatures $T>10^8$~K, and oxygen
number fractions $x_\mathrm{O}\leqslant 0.99$:
\beq
   T_\mathrm{fit}(\rho<\rho_\mathrm{lim}) = \frac{
   T_1 \,[1+(\rho_1/\rho)^{\alpha_1}]^{\alpha_2}
   }{
   \{1+[\alpha_0/\ln(\rho_\mathrm{lim}/\rho)]^{2.7}\}^{0.2}
   \,
   [1+(\rho_2/\rho)^2]^{\alpha_3}
   },
\label{ignfit}
\eeq
where $\rho_\mathrm{lim}$, $\rho_1$, $\rho_2$, $T_1$, and 
$\alpha_0$\,--\,$\alpha_3$ are fitting parameters,
given in Table~\ref{tab:ignfit}. 

\begin{table}
\caption{\label{tab:ignfit}
Parameters of \req{ignfit} for the cases 
$^{16}$O+$^{16}$O reaction in pure oxygen plasma and
$^{12}$C+$^{12}$C reaction in a mixture of carbon and oxygen with
number fractions $x_\mathrm{C}$ and $x_\mathrm{O}$, respectively
($0 \leqslant x_\mathrm{O}
\leqslant 0.99$, $x_\mathrm{C} = 1-x_\mathrm{O}$).}
\begin{tabular}{lll}
\hline\hline
Parameter & Pure $^{16}$O  & $^{12}$C+$^{12}$C in $^{12}$C/$^{16}$O
 mixture \rule[-1.2ex]{0pt}{3.8ex} \\
\hline
$\log_{10} \rho_\mathrm{lim}$ (\gcc) \rule{0pt}{2.3ex} &
  $10.6902$ &
    $9.76+0.025 x_\mathrm{O} - 0.47 \ln x_\mathrm{C}$ \\
$\log_{10} \rho_1$ (\gcc) &
  $7.2$ &
    $4.82+0.6 x_\mathrm{O} - 0.2\ln x_\mathrm{C}$ \\
$\log_{10} \rho_2$ (\gcc) &
  $6.6$ &
    $2.8+2.2 x_\mathrm{O}^4$ \\
$\log_{10} T_1$ (K) &
  $9.161$ &
    $8.75+0.015 x_\mathrm{O} - 0.033\ln x_\mathrm{C}$ \\
$\alpha_0$ &
  3.55 &
    4.2 \\
$\alpha_1$ &
  $1.4$ &
    $0.5+3 x_\mathrm{O}^4$ \\
$\alpha_2$ &
  $0.155$ &
    $0.084/\alpha_1 - 0.0029\ln x_\mathrm{C}$ \\
$\alpha_3$ &
  $0.085$ &
    $-0.0053\ln x_\mathrm{C}$ \\
\hline
\end{tabular}
\end{table}
   
At  high $\rho$, the ignition curves 
determine the
critical density of the ignition, rather than the critical
temperature, as at smaller $\rho$. Therefore, there is no sense
to measure the fit error by differences between
the model ($T_\mathrm{fit}$) and data ($T_\mathrm{dat}$) values
of temperature at a fixed $\rho$. Instead, 
we measure the fit error by the geometric
distance in the $\log\rho$\,--\,$\log(T^4)$ plane between 
numerical points ($\rho_\mathrm{dat}$,
$T_\mathrm{dat}^4$) and the line $T_\mathrm{fit}^4(\rho)$
(the fourth power of $T$ is relevant because
it is proportional to the luminosity). 
The fractional error is defined as
\beq
   \varepsilon(\rho_\mathrm{dat}) = \min_{\rho}
   \left[
    \left(
   \frac{T_\mathrm{fit}^4(\rho)-T_\mathrm{dat}^4}{T_\mathrm{dat}^4}
    \right)^2
   + \left(
   \frac{\rho-\rho_\mathrm{dat}}{\rho_\mathrm{dat}}
    \right)^2
   \right]^{1/2}.
\eeq
The maximum error $\max_{\rho_\mathrm{dat}}
\varepsilon(\rho_\mathrm{dat})$ varies from 0.06 for
$x_\mathrm{O}=0.5$ to 0.09 for $x_\mathrm{O}=0$ and 
$x_\mathrm{O}=0.99$. For oxygen, $\max_{\rho_\mathrm{dat}}
\varepsilon(\rho_\mathrm{dat})=0.05$ under the condition
that $T>1.7\times10^8$~K (at lower temperatures the ignition
curve is unreliable anyway). This is a
good accuracy, given that the considered $\rho$ and
$T^4$ values span several orders of magnitude.



\begin{thebibliography}{88}

\bibitem[Aguilera et al.(2006)]{Aguilera-ea06}
Aguilera, E.F., Rosales, P., Martinez-Quiroz, E., et al.
2006,
\prc, {73}, 064601

\bibitem[Alastuey \& Jancovici(1978)]{AlaJanco78}
Alastuey, A., \& Jancovici, B.
1978,
\apj, {226}, 1034

\bibitem[Beard et al.(2010)]{Beard-ea10}
Beard, M., Afanasjev, A.~V., Chamon, L.~C., et al.
2010,
At.\ Data Nucl.\ Data Tables, {96}, 541

\bibitem[Brown \& Bildsten(1998)]{BrownBildsten98}
Brown, E.~F.,  \& Bildsten, L.
1998, 
\apj, {496}, 915

\bibitem[Cameron(1959)]{Cameron59}
Cameron, A.~G.~W.
1959,
\apj, {130}, 916

\bibitem[Chabrier(1990)]{Chabrier90}
Chabrier, G.
1990,
J.\ Phys.\ (Paris), {51}, 1607

\bibitem[Chabrier \& Ashcroft(1990)]{ChabrierAshcroft90}
Chabrier, G., \& Ashcroft, N.~W.
1990,
\pra, {42}, 2284

\bibitem[Chabrier \& Baraffe(2000)]{ChabrierBaraffe00}
Chabrier, G., \& Baraffe, I.
2000,
\araa, {38}, 337

\bibitem[Chabrier \& Potekhin(1998)]{CP98}
Chabrier, G., \& Potekhin, A.~Y.
1998,
\pre, {58}, 4941
 
\bibitem[Chugunov \& DeWitt(2009a)]{ChugunovDW09a}
Chugunov, A.~I., \& DeWitt, H.~E.
2009a,
\prc, {80}, 014611
 
\bibitem[Chugunov \& DeWitt(2009b)]{ChugunovDW09b}
Chugunov, A.~I., \& DeWitt, H.~E.
2009b,
Contrib.\ Plasma Phys., {49}, 696
 
\bibitem[Chugunov et al.(2007)Chugunov, DeWitt, \& Yakovlev]{ChugunovDWY07}
Chugunov, A.~I., DeWitt, H.~E., \& Yakovlev, D.~G.
2007,
\prd, {76}, 025028

\bibitem[Cumming \& Bildsten(2001)]{CummingBildsten01}
Cumming, A.,  \& Bildsten, L.
2001,
 \apj, {559}, L127

\bibitem[Debye \& H\"uckel(1923)]{DH}
Debye, P., \& H\"uckel, E.
1923,
Physikalische Z., {24}, 185

\bibitem[DeWitt et al.(1973)]{DeWittGC73}
DeWitt, H.~E., Graboske, H.~C., \& Cooper, M.~S.
1973,
\apj, {181}, 439

\bibitem[Fortin et al.(2010)]{Fortin-ea10}
Fortin, M., Grill, F., Margueron, J., Page, D., \& Sandulescu, N.
2010,
\prc, {82}, 065804

\bibitem[Fowler et al.(1975)]{FowlerCZ75}
Fowler, W.~A., Caughlan, G.~R., \& Zimmerman, B.
1975,
\araa, {13}, 69
 
\bibitem[Fushiki \& Lamb(1987)]{FushikiLamb87}
Fushiki, I., \& Lamb, D.~Q.
1987,
\apj, {323}, L55

\bibitem[Galam \& Hansen(1976)]{GalamHansen76}
Galam, S., \& Hansen, J.~P.
1976,
\pra, {14}, 816

\bibitem[Gamow(1928)]{Gamow28}
Gamow, G.
1928,
Z.\ Phys., 51, 204

\bibitem[Gasques et al.(2005)]{Gasques-ea05}
Gasques, L.~R., Afanasjev, A.~V., Aguilera, E.~F., et al.
\prc, {72}, 025806

\bibitem[Gasques et al.(2007)]{Gasques_ea07}
Gasques, L.~R., Brown, E.~F., Chieffi, A., et al.
\prc, {76}, 035802

\bibitem[Gnedin et al.(2001)Gnedin, Yakovlev, \& Potekhin]{GYP}
Gnedin, O.Y., Yakovlev, D.G., \& Potekhin, A.Y.
2001,
\mnras, {324}, 725

\bibitem[Gudmundsson et al.(1983)Gudmundsson, Pethick, \& Epstein]{GPE}
Gudmundsson, E.~H., Pethick, C.~J., \& Epstein, R.~I.
1983,
\apj, {272}, 286

\bibitem[Hansen \& Van Horn(1975)]{HansenVanHorn75}
Hansen, C~J., \& Van Horn, H.~M.
1975,
\apj, {195}, 735

\bibitem[Hansen \& Vieillefosse(1976)]{HansenVieillefosse76}
Hansen, J.~P., \& Vieillefosse, P.
1976,
\prl, {37}, 391

\bibitem[Hillebrandt \& Niemeyer(2000)]{HillebrandtNiemeyer00}
Hillebrandt, W., \& Niemeyer, J.~C.
\araa, {38}, 191
 
\bibitem[Hoyle \& Fowler(1960)]{HoyleFowler60}
Hoyle, F., \& Fowler, W.~A.
1960,
\apj, {132}, 565

\bibitem[Ichimaru(1993)]{Ichimaru93}
Ichimaru, S.
1993,
Rev.\ Mod.\ Phys., {65}, 255

\bibitem[Ichimaru \& Kitamura(1996)]{IchimaruKitamura96}
Ichimaru, S., \& Kitamura, H.
1996,
\pasj, {48}, 613

\bibitem[Itoh et al.(1977)Itoh, Totsui, \& Ichimaru]{ItohTI77}
Itoh, N., Totsui, H., \& Ichimaru, S.
1977,
\apj, {218}, 477

\bibitem[Itoh et al.(1990)Itoh, Kuwashima, \& Munakata]{ItohKM90}
Itoh, N., Kuwashima, F., \& Munakata, H.
1990,
\apj, {362}, 620

\bibitem[Jancovici(1977)]{Jancovici77}
Jancovici, B.
1977,
J.\ Stat.\ Phys. {17}, 357

\bibitem[Kaminker et al.(2009)]{Kaminker-ea09}
Kaminker, A.~D., Potekhin, A.~Y., Yakovlev, D.~G., \& Chabrier, G.
2009,
\mnras, {395}, 2257

\bibitem[Kitamura(2000)]{Kitamura00}
Kitamura, H.
2000,
\apj, {539}, 888

\bibitem[Lattimer et al.(1994)]{Lattimer-ea94}
Lattimer, J.~M., Van Riper, K.~A., Prakash, M., \& Prakash, M.
1994,
\apj, {425}, 802

\bibitem[Mereghetti(2008)]{Mereghetti08}
Mereghetti, S.
2008,
\aapr, {15}, 225

\bibitem[Militzer \& Pollock(2005)]{MilitzerPollock05}
Militzer, B., \& Pollock, E.~L.
2005,
\prb, {71}, 134303

\bibitem[Mitler(1977)]{Mitler77}
Mitler, H.~E.
1977,
\apj, {212}, 513

\bibitem[Ogata(1997)]{Ogata97}
Ogata, S.
1997,
\apj, {481}, 883

\bibitem[Pollock \& Militzer(2004)]{PollockMilitzer04}
Pollock, E.~L., \& Militzer, B.
2004,
\prl, {92}, 021101

\bibitem[Potekhin(1999)]{P99}
Potekhin, A.~Y.
1999,
\aap, {351}, 787 

\bibitem[Potekhin \& Chabrier(2000)]{PC00}
Potekhin, A.~Y., \& Chabrier, G.
2000,
\pre, {62}, 8554

\bibitem[Potekhin et al.(2003)]{Potekhin-ea03}
Potekhin, A.~Y., Yakovlev, D.~G., Chabrier, G., \& Gnedin, O.~Y.
2003,
\apj, {594}, 404

\bibitem[Potekhin et al.(2009)]{Potekhin-ea09}
Potekhin, A.~Y., Chabrier, G., Chugunov, A.~I., DeWitt, H.~E.,
\& Rogers, F.~J.
2009,
\pre, {80}, 047401

\bibitem[Richardson et al.(1979)Richardson, Van Horn, \& Savedoff]{Richardson-ea79}
Richardson, M.~B., Van Horn, H.~M., \& Savedoff, M.~P.
 1979, 
\apjs, {39}, 29

\bibitem[Sahrling \& Chabrier(1998)]{SahrlingChabrier98}
Sahrling, M.,  \& Chabrier, G.
 1998, 
\apj, {493}, 879

\bibitem[Salpeter(1954)]{Salpeter54}
Salpeter, E.~E.
1954,
Australian J.~Phys., {7}, 373

\bibitem[Salpeter \& Van Horn(1969)]{SalpeterVH69}
Salpeter, E.~E., \& Van Horn, H.~M.
1969,
\apj, {155}, 183

\bibitem[Schatzman(1948)]{Schatzman48}
Schatzman, E.
1948,
J.~Phys.\ Rad., {9}, 46

\bibitem[Urpin \& Yakovlev(1980)]{UrpinYakovlev80}
Urpin, V.~A., \& Yakovlev, D.~G.
1980,
\sovast, {24}, 425

\bibitem[Yakovlev \& Shalybkov(1989)]{YaSha}
Yakovlev, D.~G., \& Shalybkov, D.~A., 
1989,
Sov.\ Sci.\ Rev., Ser.\ E: Astrophys.\ Space Phys., {7}, 311

\bibitem[Yakovlev et al.(2001)]{Yakovlev-ea01}
Yakovlev, D.G., Kaminker, A.D., Gnedin, O.Y., \& Haensel, P.
 2001,
\physrep, {354}, 1

\bibitem[Yakovlev et al.(2006)]{Yakovlev-ea06}
Yakovlev, D.~G., Gasques, L.~R., Afanasjev, A.~V., Beard, M., \& Wiescher, M.
2006,
\prc, {74}, 035803

\bibitem[Yakovlev et al.(2010)]{Yakovlev-ea10}
Yakovlev, D.~G., Beard, M., Gasques, L.~R., \& Wiescher, M.
2010,
\prc, {82}, 044609

\end{thebibliography}
\end{document}